\documentclass[12pt]{article}

\usepackage{scicite}

\usepackage{times}

\usepackage{graphicx}
\usepackage{nicefrac} 
\usepackage[hidelinks]{hyperref} 
\usepackage{graphicx}
\usepackage{color}
\usepackage{soul}
\usepackage[labelfont=bf]{caption}
\usepackage{siunitx} 
\DeclareSIUnit\Angs{\angstrom}

\newcommand{\tise}{TiSe$_2$~}
\newcommand{\ttise}{TiSe$_2$}

\newcommand{\flux}{\,\nicefrac{$\mu J$}{$cm^2$}\,}
\newcommand{\kp}{\si{\Angs}$^{\text{-}1}$}

\newcommand{\vb}{Se$_{4p\text{-}1}$~}

\newenvironment{sciabstract}{%
\begin{quote} \bf}
{\end{quote}}

\title{Ultrafast creation of a light induced semimetallic state in strongly  excited 1T-\tise} 

\author{Maximilian Huber $^{1}$, Yi Lin $^{1,2}$, Giovanni Marini$^{3,4}$, Luca Moreschini$^{1}$, \\ Chris Jozwiak$^{5}$,  Aaron Bostwick$^{5}$,  Matteo Calandra$^{3,4,6}$ \\ and Alessandra Lanzara$^{1,7,8\ast}$\\
\\
\normalsize{$^{1}$Materials Science Division, Lawrence Berkeley National Laboratory,}\\
\normalsize{Berkeley, CA 94720, USA}\\
\normalsize{$^{2}$Department of Physics and Astronomy, University of Alabama, Tuscaloosa AL 35487, USA}\\
\normalsize{$^{3}$Graphene Labs, Fondazione Istituto Italiano di Tecnologia,} \\ 
\normalsize{I-16163 Genova, Italy}\\
\normalsize{$^{4}$Department of Physics, University of Trento, 38123 Povo,
Italy}\\
\normalsize{$^{5}$Advanced Light Source, Lawrence Berkeley National Laboratory, } \\ 
\normalsize{Berkeley, California 94720, USA}\\
\normalsize{$^{6}$Sorbonne Universit\'e, CNRS, Institut des Nanosciences de Paris, F-75252 Paris, France} \\
\normalsize{$^{7}$Physics Department, University of California Berkeley, Berkeley, CA 94720, USA}\\
\normalsize{$^{8}$Kavli Energy NanoScience Institute, Berkeley, CA 94720, USA}\\
\normalsize{$^\ast$To whom correspondence should be addressed; E-mail:  alanzara@lbl.gov}
}

\date{}

\begin{document} 


\baselineskip24pt


\maketitle 

\begin{sciabstract}
Screening, a ubiquitous phenomenon in condensed matter physics associated with the shielding of electric fields in a solid by surrounding charges has been widely adopted as a mean to modify a material's electrical, optical and transport properties. While so far most of previous studies have relied on static changes of screening through doping or gating, in this study we demonstrate that screening can also drive the onset of new quantum states on the ultrafast timescale.  By using time and angle resolved photoemission spectroscopy we study the response of 1T-\ttise, a prototypical charge density wave material, to intense laser pulses. Specifically we show that within the strong excitation regime 1T-\tise evolves almost instantly from a gapped charge density wave state into a semimetallic state. These changes in band structure are fluence and time dependent, leading to a substantial indirect band overlap on the order of $\sim$350\,meV for the highest fluence studied. This drastic response, featuring a substantial change of the valence band shape from parabolic to linearly dispersing across the Fermi level, is beyond the more standard screening-induced renormalization of band structure and collective excitations, widely studied in the literature. By systematically comparing the changes in bandstructure over time and excitation strength with theoretical calculations we find that the appearance of the semimetallic state is likely caused by a dramatic reduction of the screening length, not only affecting electron-electron, but possibly also the electron-lattice interactions. In summary, this work showcases how optical excitation enables the screening driven design of a non-equilibrium semimetallic phase in \ttise, possibly providing a general pathway into highly screened phases in other strongly correlated materials.

\end{sciabstract}


Understanding the effect of screening is a prerequisite step for understanding and manipulating many body interactions in quantum materials in and out of equilibrium. Most previous studies have focused on manipulating screening properties by tuning the carrier density (doping\,\cite{Siegel2011, Siegel2013, Berggren1981, Riis2020, Ye2014} or electrostatic gating\,\cite{Dale2022, Efetov2010, Elias2011}). More recently, sample thickness has been utilized as another control knob for instance in transition metal dichalcogenides (TMDCs), where the reduced dimensionality in the monolayer limit leads to reduced screening and thus to drastically enhanced exciton binding energies\,\cite{Ugeda2014}. Photodoping is a particularly effective tool to manipulate the electronic properties of a material via screening compared to doping with external carriers in equilibrium, as the optical excitation leads to the quasi-instantaneous creation of hot electron and holes simultaneously. Indeed, the photoinduced renormalization of band gap or many body interactions have been reported in, among others, semiconductors\,\cite{Puppin2022, Lin2022}, superconductors\,\cite{Smallwood2012, Smallwood2014, Zhang2014} and topological insulators\,\cite{Sam2023}. In extreme cases, such abrupt changes can drive materials into non-equilibrium phases that cannot be obtained under equilibrium conditions\,\cite{Torre2021, Stojchevska2014, Kogar2020, Fausti2011, Duan2023}. The sensitivity of the prototypical charge density wave (CDW) material 1T-\tise to electronic interactions\,\cite{Hellgren2017, Hellgren2021} makes it an ideal system to study the effect of photoinduced screening-driven states beyond equilibrium. Especially the strong Coulomb repulsion due to the localized nature of the Ti$_{3d}$ orbitals makes the theoretical description of \tise intricate even at equilibrium and as a result (semi)local functionals are not able to accurately describe the electronic structure of the normal and CDW state\,\cite{Hellgren2017}. While the introduction of the Hubbard\,U can in principle cure this well known delocalization error, leading to very accurate electronic structures, it fails to reproduce the structural properties and the CDW instability\,\cite{Bianco2015}. On the other hand, the use of the computationally more expensive hybrid functionals allows to calculate both the electronic structure and phonon spectrum in accordance with experimental data\,\cite{Hellgren2017, Hellgren2021}. Studying \tise under highly out-of-equilibrium conditions is furthermore of interest in view of the recent report of a metastable metallic phase\,\cite{Duan2023}, and to shed a new light into one of the ongoing debates in the field, i.e.\ how its electronic structure evolves from the CDW to the normal state\,\cite{Rossnagel2002, Rossnagel2010, Rossnagel2011, Rohwer2011, Porer2014a, Wegner2018}.  \\ 
In this regard angle resolved photoemission spectroscopy (ARPES) is the ideal tool to study screening effects in and out of equilibrium\,\cite{Gatti2020, Siegel2011, Siegel2013}, as the simultaneous momentum and energy resolution allow to directly visualize the evolution of the band structure upon photoexcitation as well as detect changes of many-body interactions\,\cite{Damascelli2003, Iwasawa$_2$020}, information not easily directly accessible through other techniques. 
Our results reveal an ultrafast transition from the gapped CDW phase into a semimetallic state, which happens within the temporal resolution of our experiment, i.e.\ in under 55\,fs. Careful analysis as a function of excitation density shows that the band structure gets increasingly renormalized with increasing pump fluence, with the valence band gradually opening up until it shows an almost linear dispersion crossing the Fermi level. With the band structure being clearly different than both the equilibrium high and low temperature state, we utilize density functional theory (DFT) simulations to show an remarkably similarity between the experimental data and calculated quasiparticle band structures for the CDW state with screened electron-electron interactions and substantially reduced lattice order. In summary this work demonstrates how intense laser pulse can drive non-equilibrium states of matter where electron-electron and possibly electron-lattice interactions are highly suppressed.

\section*{Results}

\begin{figure} [h]
	\centering
	\includegraphics[width=0.95\textwidth]{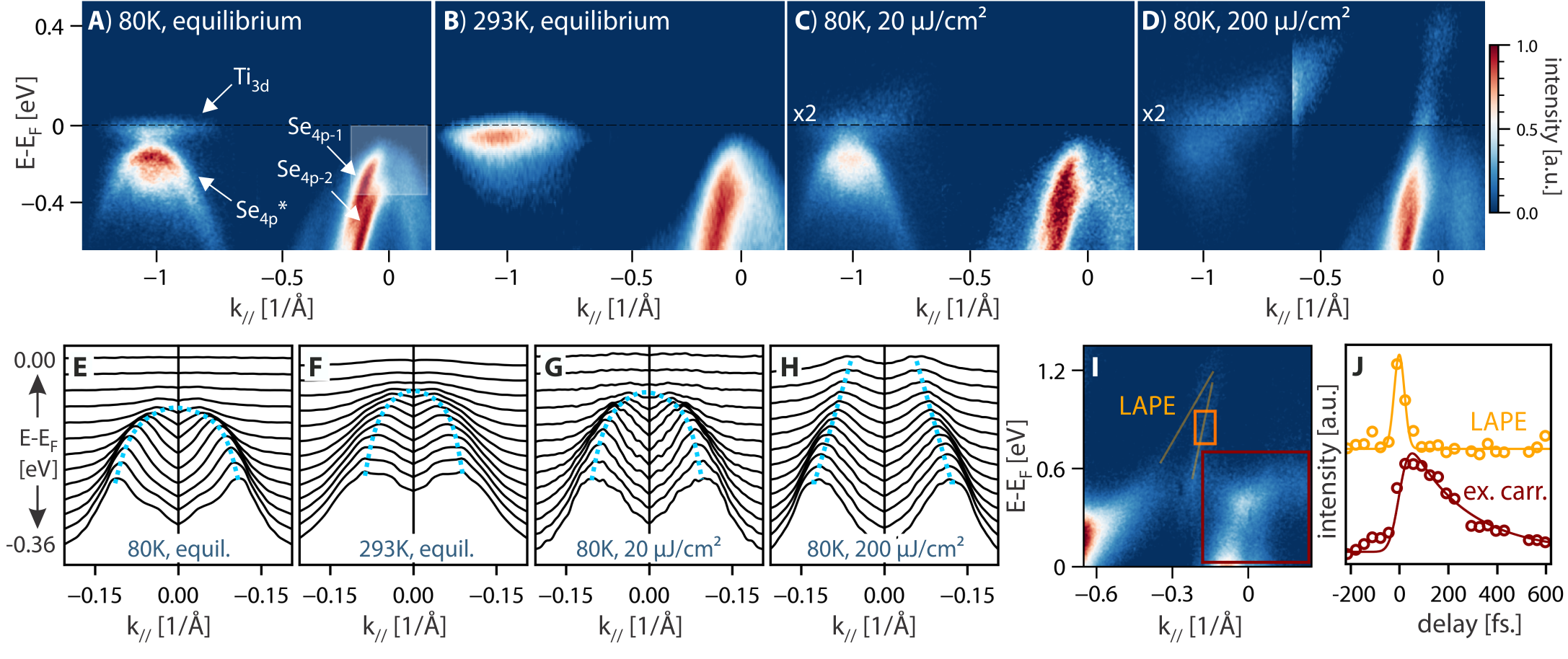}
	\caption{\textbf{Comparison of the electronic structure of \tise under different conditions.} a-b) ARPES spectra of the $L$ and $A$-point in equilibrium at 80\,K (a) and at room temperature (b), respectively.  c-d) ARPES spectra after excitation with 780\,nm pump pulses with 20\,\flux (panel\,c) and 160 ($L$ point) and 200\,\flux ($A$ point) (panel\,d), respectively, after 80\,fs. Spectra at $L$ and $A$ point are normalized independently. For clarity excited states in panel\,c and d are shown with higher intensity and after subtraction of an exponential background above the Fermi level.  The Fermi level is indicated by the dashed black line. g-j) Stacked momentum distribution curves taken between -0.36 and 0\,eV with focus on the $A$ point corresponding to spectra in panels a-d. Band dispersion is schematically shown as guide to the eye by the dashed blue lines. The region from which MDCs are extracted is shown in panel\,a by the white shaded area.  i) APRES image plot at $A$ point after excitation with 180\,\flux around 0\,fs.  Bands due to the LAPE effect are marked in orange. For clarity a exponential background was subtracted.  j) Dynamics of LAPE (orange) and excited carriers (red) after excitation with 180\,\flux. Integration regions are indicated in panel\,i. Solid lines are fits to the data points (Gaussian function for the LAPE, exponential decay for excited carriers).}
	\label{fig:fig1}
	\end{figure}

\autoref{fig:fig1} shows the evolution of the electronic band structure of the CDW phase following an optical excitation by a near-infrared pump pulse of 1.6\,eV (780\,nm). Data are taken with a probe energy of 22.3\,eV. See Methods section of the SM for more details of the extreme-ultraviolet (XUV) tr-ARPES experiment and setup. Panel\,a shows the experimental equilibrium quasiparticle band structure measured at 80\,K, i.e.\ below the CDW transition temperature (T$_{CDW} \sim$ 200\,K\,\cite{DiSalvo1976}), in the $k_z$ plane close to the $A$-$L$ direction\,\cite{Rossnagel2002, Watson2019, Chen2016b} in the bulk Brillouin zone notation. The onset of the CDW state in \tise is accompanied by a periodic lattice distortion (PLD), which leads to a doubling of the unit cell size into a (2$\times$2$\times$2) superstructure (i.e.\ a folding of the BZ) from the typical (1$\times$1$\times$1) structure in the normal state\,\cite{DiSalvo1976, Rossnagel2002}. As a consequence of CDW/PLD formation the Se$_{4p}$ valence band from the $A$ point gets folded onto the $L$ point (Se$_{4p}$* band), where a gap between the Se$_{4p}$* band and the Ti$_{3d}$ conduction band opens up. The high momentum and energy resolution of this experiment allow us to directly visualize those spectral fingerprints of the CDW state. Furthermore, the spin-orbit splitting of the valence band (Se$_{4p\text{-}1}$ and Se$_{4p\text{-}2}$) is clearly resolved. The experimental band structure of the CDW state is in overall very good agreement with our band structure calculations in \autoref{fig:fig4}a and previous reports\,\cite{Rossnagel2010, Rohwer2011, Watson2019}.
At room temperature (panel\,b), i.e.\ above the CDW transition temperature (T$_{CDW} \sim$200\,K\,\cite{DiSalvo1976})  the folded band at $L$ disappears almost completely and the valence band at $A$ shifts slightly upwards (by $\sim50$\,meV), still leaving a $\sim$80\,meV indirect gap between the valence and conduction band, in accordance with previous reports\,\cite{Rossnagel2002, Watson2019, Monney2010a, Huber123}. \autoref{fig:fig1}\,c and\,d show the evolution of the CDW state following pump excitation, in the weakly and strongly excited regimes, respectively. In the weakly excited regime (panel\,c) we observe a reduction of the intensity of the backfolded band, a slight decrease of the CDW gap and the occupation of the Ti$_{3d}$ band by hot electrons, in line with previous reports\,\cite{Huber123, Huber2022, Mathias2016, Rohwer2011}. Surprisingly, when the system is excited with stronger laser pulses (panel\,d) a completely new feature appears at the $A$ point. The valence band opens up with its dispersion changing drastically from the parabolic to a linearly dispersive band shape, which extends all the way above the Fermi level. Such a state has to the best of our knowledge never been reported before, neither in previous time resolved ARPES studies\,\cite{Rohwer2011, Mathias2016, Hellmann2012, Duan2021, Hedayat2019a, Duan2023} nor in equilibrium experiments at high temperature or with doping\,\cite{Rossnagel2010, Rossnagel2002}. To further visualize the light induced dispersion changes, in panels e-h we show the momentum distribution curves (MDCs) (curves at constant energy as a function of momentum) with focus on the valence band at the $A$ point. Weak excitation or raising the temperature in equilibrium (compare panels\,f and\,g with panel\,e) induces a small upshift of the Se$_{4p}$ bands and slightly affects the effective mass in comparison to the CDW equilibrium phase in\,\autoref{fig:fig1}e. Despite the minor changes, the overall shape of the band remains parabolic in both cases, with the maximum below the Fermi level. For further details on the low excitation regime we refer to our previous work\,\cite{Huber2022, Huber123}, where we have extensively studied photoinduced changes of both the dispersion and bandgap. In drastic contrast, in the strong excitation regime the low temperature spectrum is remarkably different from the equilibrium low and high temperature spectra (compare panels e-h). Specifically the data shows the opening of the \vb band with consequent crossing of the Fermi level and an evolution from a parabolic to an almost linear dispersion. \\
After showing the existence of a light induced semimetallic state after strong excitation we want to turn the attention to the ultrafast timescale on which it emerges. \autoref{fig:fig1}\,i shows the 2D image plot of the excited states above the Fermi level taken with very high statistics. The excellent energy resolution allows to resolve, in addition to the linear excited state, the two Se$_{4p\text{-}1}$  and Se$_{4p-2}$ replica bands due to the laser-assisted photoelectric effect (LAPE, indicated by orange lines). 
The LAPE effect is a dressing of the photoemitted electron with the pump laser field\,\cite{Saathoff2008} and hence it can only occur within the temporal width of the pump pulse.  Therefore, its coexistence with the semimetallic state suggests that the transition into this new regime happens almost instantly after excitation within the temporal resolution of the experiment. Panel\,j reports the dynamics of the excited carriers above the $A$ point (integration region shown by red box in panel\,i). The orange curve shows the dynamics of the LAPE effect which as mentioned above by its own nature occurs at t=0, and is thus often utilized in pump and probe ARPES experiments as reference for the temporal overlap. Fitting the curve with a Gaussian gives a full-width-half-maximum of $55 \pm 7$\,fs, which indicates the temporal resolution of the experiment. In contrast to the LAPE effect, the excited carrier population (red) reaches its maximum shortly after excitation at approximately t=60\,fs, indicating that the observed semimetallic state is not an intrinsic effect due to the photoemission process. By fitting the data with an exponential, convoluted with a Gaussian (to account for the experimental time resolution), we extract a decay time of 210\,fs$\pm$17. This fast decay of excited carriers is a further sign of a light-induced metallization of the sample and indicates that there are no larger gaps. In the presence of a pronounced gap we would expect to observe a bottleneck in the carrier relaxation as it is the case for other gapped systems\,\cite{Smallwood2015, Mori2023}. \\

\begin{figure} [h]
	\centering
	\includegraphics[width=0.95\textwidth]{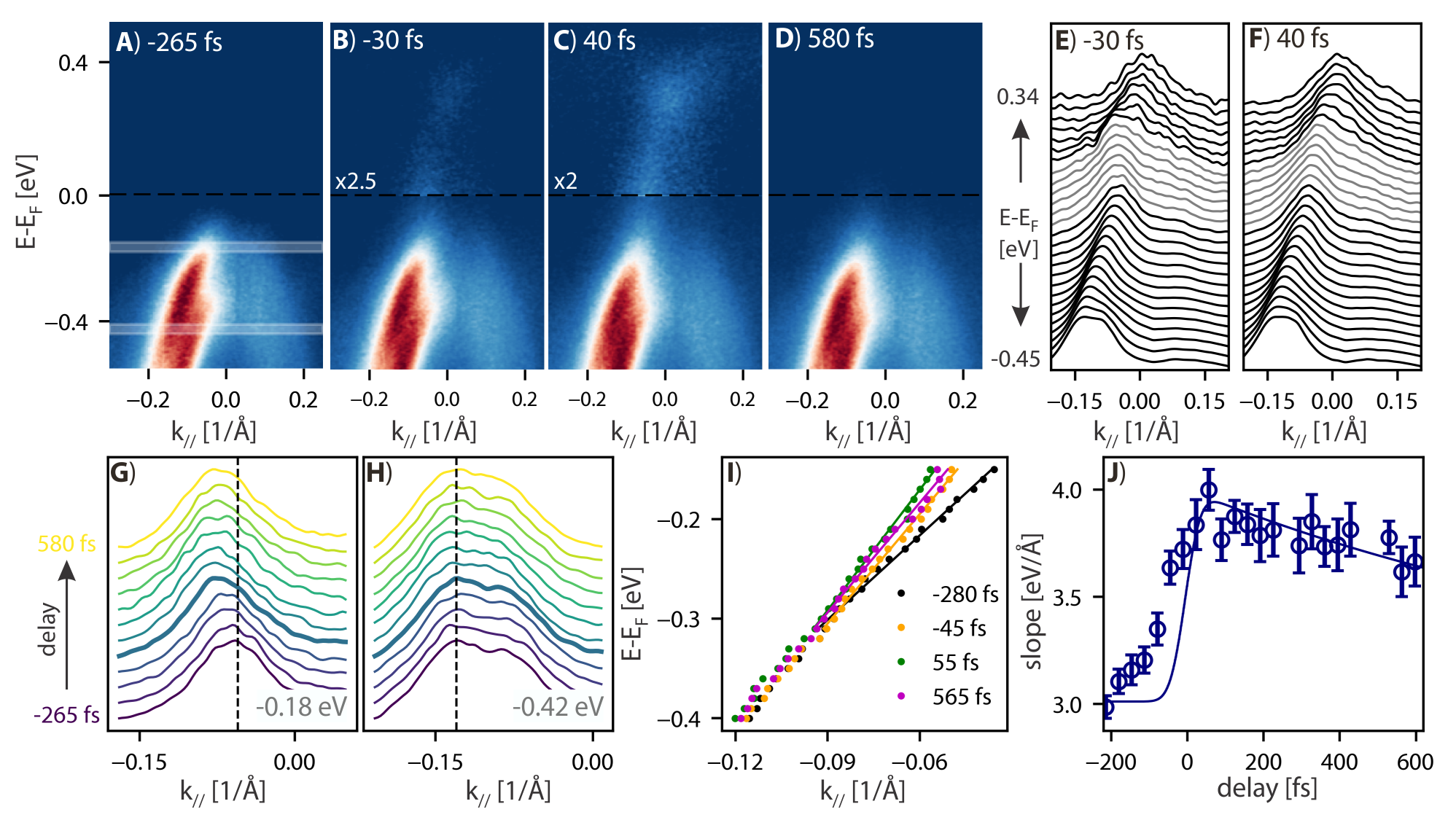}
	\caption{\textbf{Time dependence of the semimetallic state.} a-d) ARPES image plots at delay of -265 (a), -30 (b), 40 (c) and 580\,fs (d) after excitation with 180\,\flux. All spectra were aligned in energy at to the side of the Se$_{4p\text{-}1}$ band at -0.12\kp.  For better clarity the shown plots are averaged over 66\,fs (i.e.\ two delay values).  e-f) Stacked MDCs corresponding to the spectra in panel\,b and\,c, respectively. For clarity each MDC slice is normalized to the same height. The energy region in which the band is less dispersive is indicated by the grey lines.  g-h) MDCs at different delay values integrated around -0.18\,eV (panel\,g) and -0.42\,eV (panel\,h), respectively. Integration regions are schematically shown in panel\,a. For enhanced clarity the MDCs were averaged over 66\,fs. i) Extracted dispersion of the Se$_{4p\text{-}1}$ band for different delays with focus on below E$_F$. Solid lines are linear fits to the data points.  j) Dynamical behavior of the slope extracted from linear fits to the valence band dispersion.}
	\label{fig:fig2}
	\end{figure}

\begin{figure} 
	\centering
	\includegraphics[width=0.95\textwidth]{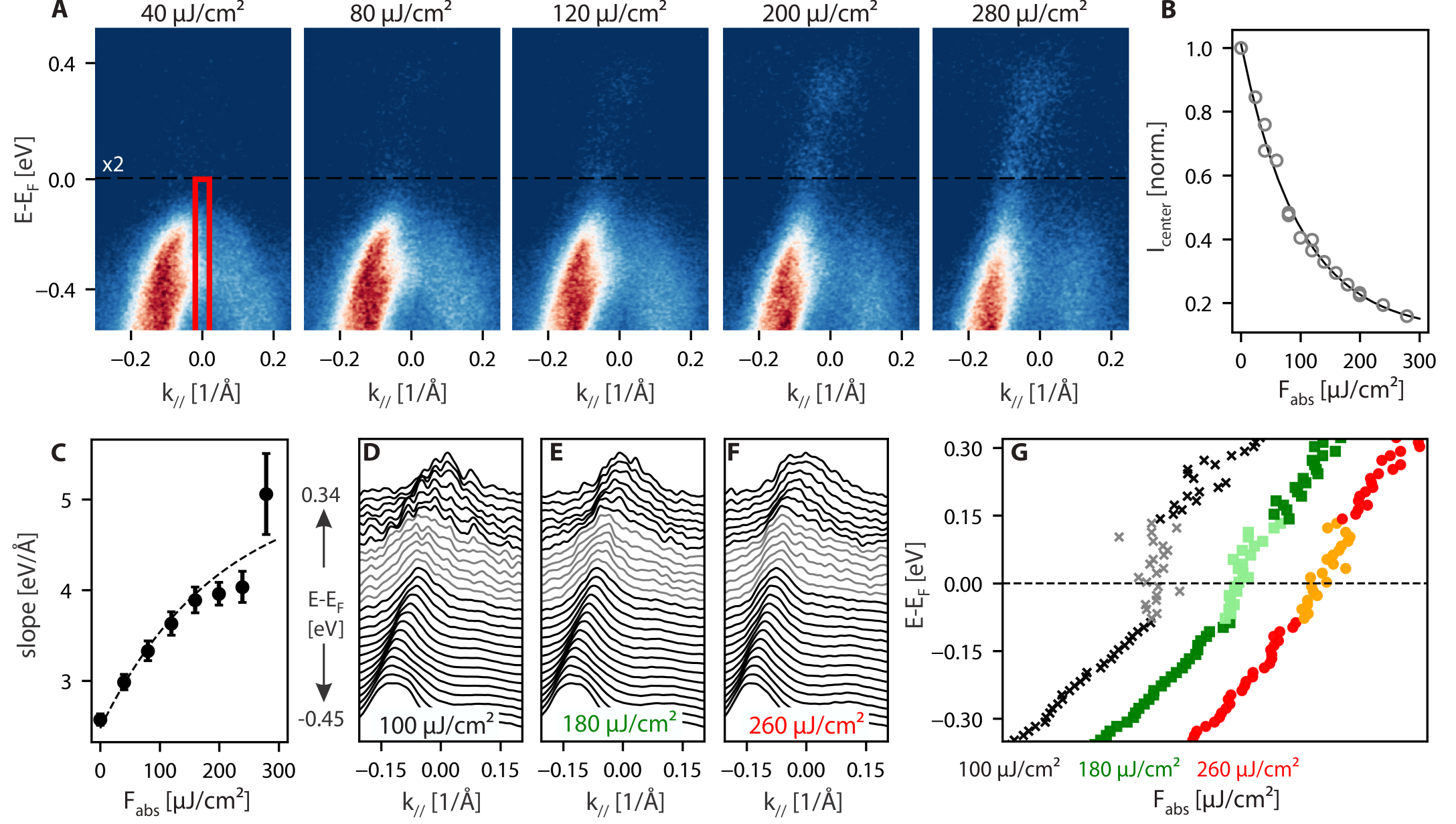}
	\caption{\textbf{Fluence dependence of the semimetallic state.} a) ARPES plot at a delay of 85\,fs after excitation with various fluences. b) Integrated intensity at $A$ point. Integration region is shown by the red box in panel\,a. c) Extracted slope of the fitted Se$_{4p\text{-}1}$ dispersion for different fluence values. d-f) Stacked normalized MDCs for different fluences. For better signal to noise ratio the shown MDCs are taken from spectra averaged over two fluence values, i.e.\ over a window of 80\,\flux after correcting for band shifts. The region in which the band is less dispersive is indicated in grey. g) Extracted dispersion of the Se$_{4p\text{-}1}$ band extracted from MDCs shown in panels d-f. }
	\label{fig:fig3}
	\end{figure}

Having shown the ultrafast and intrinsic nature of the optically driven semimetallic state, we now proceed to a quantitative study of its evolution as a function of time (\autoref{fig:fig2}) and pump fluence (\autoref{fig:fig3}). In order to compensate for rigid energy shifts for instance due to space charge effects all spectra that are analyzed in the following are aligned in energy with respect to the side of the Se$_{4p\text{-}1}$ valence bands (see Supplementary Note\,1 for more details).
\autoref{fig:fig2}\,a-d show the ARPES spectra for different delay times, where the transition from the gapped into the semimetallic state can be clearly followed. Due to the finite width of the laser pulses, carriers get excited even before the maximum of pump and probe pulse overlap at t=0\,fs. Indeed, already at very early delay times (-30\,fs) we can observe excited carriers that populate a seemingly linearly dispersing band above the Fermi level. A closer look reveals that there is a discontinuity between the dispersion of the bands populated by hot electrons above the Fermi level and the valence band below, whereas at later times (t=40\,fs, panel\,c) both bands are connected. To analyze these subtle differences in more detail, in panel\,e and\,f we plot stacked MDCs, corresponding to the spectra shown in panel\,b and\,c, respectively. To compensate for the large intensity difference of states above and below E$_F$ we normalized each MDC slice individually to the same height. At -30\,fs (panel\,e) one can clearly identify a region of about $\sim200$\,meV marked in grey where the MDC peaks do not disperse. The presence of non-dispersive MDCs in ARPES spectra hints towards an energy gap\,\cite{Zhou2007, Zhou2008}, where in this particular case the persistence of peaks in the gap region is caused by broadening effects due to the highly out-of-equilibrium nature of the experiment and the large intensity difference between the original valence band and states above the Fermi level only populated by excited carriers. In contrast, at a slightly later delay time (panel\,f) the MDCs peaks show a seemingly linear trend over the entire energy range, indicating an almost continuous, linearly dispersing band with no or only a small gap. //
To better understand the opening dynamics of the valence band, in panels\,g and\,h we show the MDC spectra near the top of the valence band around -0.18\,eV and at lower energy -0.42\,eV. The spectra are integrated over an energy window of 20\,meV. Note again that in order to analyze changes in the band dispersion free of rigid band shifts the spectra were aligned in energy before extracting the MDCs (Supplementary Note\,1). In the vicinity of the Fermi level (panel\,g) the MDC peaks show a pronounced shift away from the $A$ point, with the maximum shift observed at around t=0\,fs (thick blue line). In contrast the MDCs at lower binding energies (panel\,h) barely show any shift, suggesting that the changes in the proximity of the Fermi level are in fact due to an opening of the \vb band, eventually leading to the almost linear dispersion relation almost instantly after excitation. To quantify those changes, in panel i) we show the Se$_{4p\text{-}1}$ dispersion extracted from standard Lorentzian fitting of the MDC spectra of the upper part of the valence band (see Supplementary Note\,2). We note that as shown in \autoref{fig:fig1} the band structures between equilibrium and semimetallic state are drastically different. While energy distribution curves (EDCs) are better in capturing the parabolic shape in equilibrium, MDCs are better suited to describe the dispersion of the linear bands of the semimetallic state. As we are interested in the timescale of the transformation into the semimetallic state we hence consistently extract the dispersion for all delay values using MDCs. The timescale associated with the valence band opening is shown in panel\,j, where the change of the band velocity (slope) obtained through a linear fitting of the top of the valence band is shown as a function of delay time (fits for selected delay times are shown in solid lines panel\,i). The maximum change occurs around t=0\,fs and is followed by a slow recovery. Within 600\,fs the slope has still not fully recovered back to its equilibrium value, which is in contrast to the excited carriers which decay back to equilibrium within a short time scale (\autoref{fig:fig1}j). Further discussion on the possible origin of this slow recovery is presented later.


After having shown the light induced ultrafast temporal evolution from the gapped CDW into the semimetallic state we now proceed to analyze the transition as a function of fluence. \autoref{fig:fig3}\,a shows ARPES image plots at a delay of 85\,fs for different excitation densities. At lower fluences the valence band shows the standard parabolic dispersion, typical of the CDW phase.  As the excitation density increases, the \vb band opens up and a linear dispersion relation develops, which is as discussed above signature of the onset of a semimetallic phase. To quantify the photoinduced changes we plot the ratio of the integrated intensity at the center of the $A$ point with respect to the equilibrium value as a function of fluence (panel\,b). The integration region is shown by the red box in panel\,a. After a fast initial decrease for low excitation densities, the quench of the intensity starts to saturate at higher fluences, which might be caused by a saturation of the optical absorption\,\cite{Mor2017}. As multiple factors like band depopulation or band renormalization can cause the observed decrease of intensity we report the fluence dependence of the Se$_{4p\text{-}1}$ band velocity as a more direct measure of the opening of the valence band in panel\,c. The velocity (slope) is extracted in the same way as discussed in the previous figure (\autoref{fig:fig2}). The trend can be described by a exponential function, with a fast increase of the slope at low fluences followed by a slowing down at higher fluences. For a more detailed analysis in panels d-f we plot again normalized stacked MDCs in the same way as in \autoref{fig:fig2}e-f for different fluences. By fitting the MDCs with the standard Lorentzian fitting method\,\cite{Iwasawa$_2$020, Damascelli2003}  (see Supplementary Note\,2) we extract the \vb dispersion shown in panel\,g. Similar to the data point at very early delay time (\autoref{fig:fig2}e), for weak excitation density we can identify a large region around the Fermi level where the MDC peak does not disperse, indicating a pronounced gap. With increasing fluence one can observe how this region gradually becomes more dispersive until finally at high fluences the band above and below merge into one continuous band. For the highest fluence studied (280\,\flux) we estimate an indirect band overlap between the bottom of the Ti$_{3d}$ conduction band at $L$ point and the top of the valence band at $A$ point of about $\sim$350\,meV (see Supplementary Note\,3). This value is drastically different than reported in previous equilibrium studies at room temperature, where usually a indirect gap of $\sim$75\,meV\,\cite{Rossnagel2002, Monney2010a, Watson2019} is observed and further highlights the non-equilibrium nature of the light induced semimetallic state.


\section*{Discussion}
In summary, the data shown so far unambiguously prove the existence of a linearly dispersing, photoinduced semimetallic state in \ttise, which to the best of our knowledge has not been observed so far. This might be the result of photoemission matrix element effects due to the use of different polarization\,\cite{Mathias2016} or photon energy\,\cite{Rohwer2011, Hellmann2012, Hedayat2019a, Duan2021, Duan2023} of the probe beam in previous work. As shown in Supplementary Notes\,4, the visibility of the Se$_{4p\text{-}1}$ band forming the semimetallic state is strongly dependent on the polarization of the probe beam. Moreover, we find that extrinsic factors such as sample and cleave quality tend to strongly suppress the intensity of the Se$_{4p\text{-}1}$ band. The light induced semimetallic state is remarkably different from both the equilibrium low and high temperature state (\autoref{fig:fig1}a,b and \cite{Watson2019,Rossnagel2002, Monney2010a}). Even increasing the temperature far above room temperature (Supplementary Note\,5) or heavy doping\,\cite{Mottas2019a, Rossnagel2010, Zhao2007, Adam2022, Jaouen2023} do not show the drastic renormalization of the valence band observed in this work.  Since in synchrotron experiments photon energies and polarization are easily tunable and the resolution is generally higher than for XUV-trARPES experiments, we take this as further support of the non-equilibrium nature of the light induced semimetallic state. Interestingly, \autoref{fig:fig1}d seems to suggest that there is a regime in which the band structure can become semimetallic while there is still a finite PLD (indicated by residual intensity of the folded Se$_{4p}$* band), a scenario which is impossible for both a Peierls and an excitonic insulator driven CDW phase under equilibrium conditions\,\cite{Rossnagel2011, Gruener}.

\begin{figure} 
	\centering
	\includegraphics[width=0.95\textwidth]{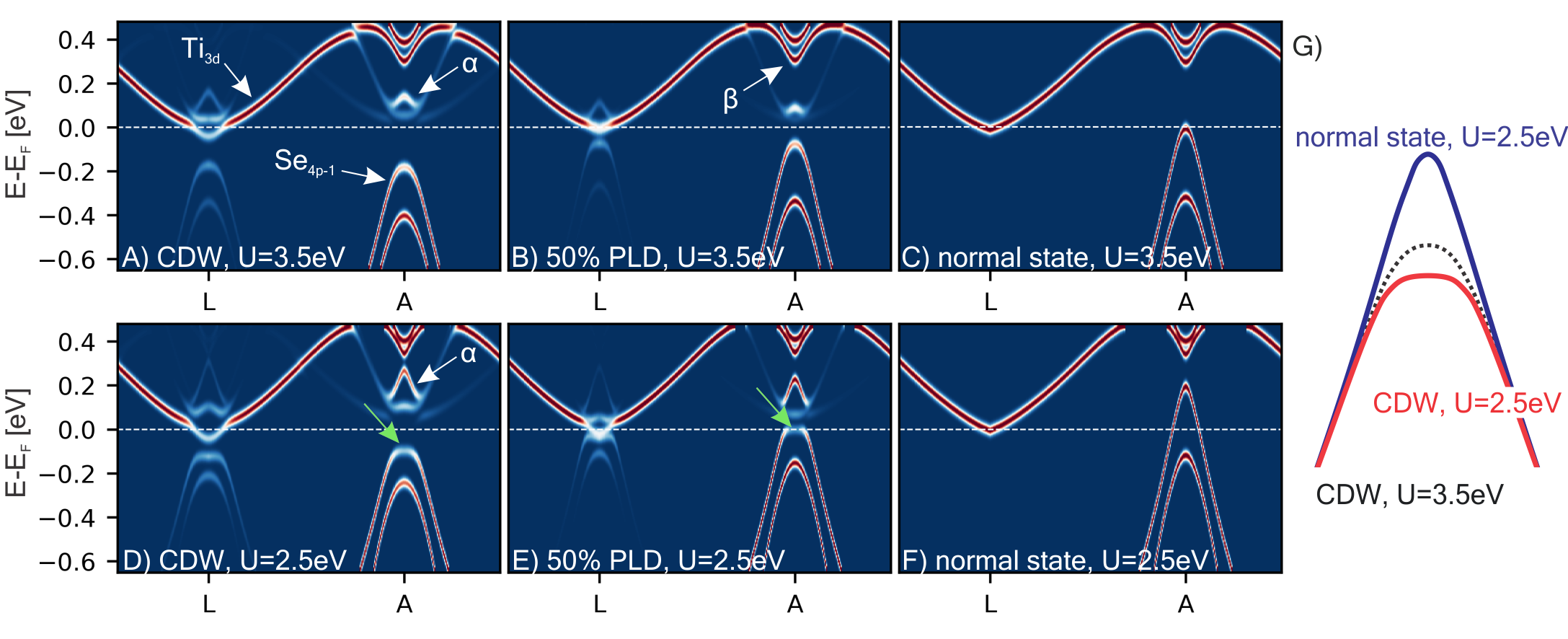}
	\caption{\textbf{Theoretical Model.} Calculated single-particle band structures for different lattice distortions and Hubbard\,U terms.}
	\label{fig:fig4}
	\end{figure}

To get a better insight in the origin of this light induced semimetallic state we perform first principles density-functional theory (DFT) calculations. Femtoseconds after laser irradiation we expect a substantial enhancement of the electronic screening of the Coulomb interaction due to the higher density of non-equilibrium free electronic carriers. In order to qualitatively capture this effect in \ttise, we describe the electronic properties of both the normal and CDW phases within the DFT+U formulation\,\cite{Bianco2015,PhysRevB.52.R5467}. Within the DFT+U approach, an enhanced screening due to free carriers naturally leads to an effective renormalization of the Hubbard\,U value\,\cite{Tancogne-Dejean2018, Tancogne-Dejean2020, Baykusheva2022}, which represents the short-range Coulomb repulsion. Besides screening of electron-electron interactions, diffraction experiments also indicate a fast suppression of PLD order\,\cite{Duan2021, Mohr-Vorobeva2011}. Thus, in order to qualitatively understand the non-equilibrium behavior of \tise in our calculations, we systematically vary the strength of lattice distortion and electron-electron interactions. This is shown in \autoref{fig:fig4}, where the electronic band structure for U$=3.5$ eV (the equilibrium\,U value) and U$=2.5$\,eV (the reduced\,U value) are compared (see Supplementary Note\,6 for further technical details on the theoretical spectra). As the system is initially found in the CDW phase, we start by analyzing the electronic structure of the CDW phase unfolded onto the Brillouin zone of the TiSe$_2$ normal phase in panel\,a, where the main features are consistent with the present experiment (compare to \autoref{fig:fig1}a) and previous literature\cite{Bianco2015}. In the upper row (panels\,a-c) we show the evolution of the band structure with decreasing PLD, which resembles the equilibrium phase transition with increasing temperature as studied in literature. The main changes at the $A$ point are: an upward shift of the valence band with respect to the CDW state (panel\,a) accompanied by a simultaneous downshift of the hole like band ($\alpha$) above the Fermi level, where both bands eventually merge into each other into a semimetallic band structure in the normal state with no PLD. At the $L$ point the main change is the disappearance of the backfolded Se bands and the closure of the CDW gap. In the bottom row we consider the effect of the enhanced screening (reduced\,U value) on the spectra. These calculations model the (possible) partial nonthermal melting of the charge-density wave order. For all lattice orders we observe a substantially modified quasiparticle dispersion, with important effects on the spectral intensity of both the valence and conduction bands in proximity of the $A$ point of the BZ. In particular, in the CDW state with decreased\,U the top of the Se$_{4p\text{-}1}$ valence band, originally parabolic, becomes flatter and loses intensity (see green arrow) while the hole like $\alpha$ band above the Fermi level acquires additional spectral weight. At the same time, the Se$_{4p\text{-}1}$ band shifts upwards and the CDW gap at $L$ point becomes smaller. Moreover, the decrease of\,U causes the $\alpha$ band to disperse across a larger energy region (with the top of the band now lying around 300\,meV above E$_F$) while still remaining clearly separated from the valence band by a gap $>200$\,meV. With decreasing PLD (panel\,e) the gap at the $A$ point progressively vanishes and the upper valence band reaches the Fermi level, together with a further decrease of the intensity of the top of the Se band at the $A$ point (green arrow). An interesting point is that while for the CDW state with unperturbed lattice the system stays semiconducting both with full and reduced Hubbard\,U (panels\,a and\,d), for $50 \%$ PLD a decrease of Hubbard\,U changes the electronic properties from semiconductor (panel\,b) to semimetal (panel\,e). Finally, for the normal state with decreased\,U, where the PLD is fully quenched into the normal 1$\times$1$\times$1 lattice order (panel\,f), we observe the merging of the hole like feature above the Fermi level ($\alpha$) with the valence band below, following the full opening of the valence band as well as a sharpening of the top. Here the overlap between valence band at $A$ and conduction band at $L$ is drastically enhanced compared to the normal state with U=3.5\,eV (panel\,c). To better visualize the changes in band structure, in panel\,g we plot the calculated Se$_{4p\text{-}1}$ valence band dispersion for different scenarios on top of each other and aligned in energy to the side of the band, identical to the alignment of our experimental data in \autoref{fig:fig2} and \autoref{fig:fig3}. The comparison highlights how in the CDW state a change in\,U only leads to a flattening of the band, whereas a decrease in PLD is necessary to cause an opening of the top of the Se band. \\
Comparing these calculations with the experimental results we find an excellent agreement between the band structures with reduced Hubbard\,U (\autoref{fig:fig4}d-f) and the ARPES spectra taken after optical excitation. For very early delays (-30\,fs, \autoref{fig:fig2}b and\,e) and in the intermediate to low fluence regime (\autoref{fig:fig3}d and\,e) we observe states where the top of the Se$_{4p\text{-}1}$ valence band partially opens up while excited carriers populate a linearly dispersing band above the Fermi level, with both bands being separated by a large energy gap on the order of $\sim 200$\,meV. These spectra resemble the calculated band structures with enhanced Coulomb screening (reduced\,U) and slightly reduced PLD (\autoref{fig:fig4}d,e). With increasing fluence, this gap closes and we observe  the opening of the parabolic valence band and its evolution into a linearly dispersing band with maximum far above the Fermi level at the $A$ point. This is similar to the calculated evolution of the band structure with decreasing PLD into the normal state with reduced\,U in \autoref{fig:fig4}e,f.  We also note that other general features of the calculations like the extension of the conduction band from $L$ point into the electron pocket at $A$ point (labeled $\beta$ in \autoref{fig:fig4}) are reproduced as well (\autoref{fig:fig1}\,i). The excellent agreement of the calculated band structures with reduced\,U and our experimental data suggests that the observed semimetallic states are driven by screening effects. This is further supported by the ultrafast timescale on which the semimetallic state appears (\autoref{fig:fig1}i,j and \autoref{fig:fig2}) and the correspondence of the opening dynamics with the build-up time of the excited carriers (compare \autoref{fig:fig2}j with \autoref{fig:fig1}j). In this context the slow recovery of the slope shown in \autoref{fig:fig2}j, contrasting the fast decay of excited carriers (\autoref{fig:fig1}j), might seem somewhat surprising. Our calculations can account for this trend under the hypothesis in which the system relaxes from the normal state with reduced\,U into a state that is more similar to the equilibrium normal state (i.e.\ with U$\sim$3.5\,eV) after 600\,fs\,\cite{Wegkamp2014}. This could eventually be also compatible with the light induced metastable state recently reported in Ref.\,\cite{Duan2023}, for which lifetimes longer than several picoseconds were observed. \\
We finally want to emphasize that according to the calculations for the opening of the top valence band a quench of PLD order is necessary (\autoref{fig:fig4}g) and for a complete transition into the semimetallic state as observed in the high fluence regime the lattice needs to transition completely into the normal state order. Since ARPES is only directly sensitive to the electronic order, it is ambiguous whether there is actually a quench of the PLD on such a fast timescale or if the semimetallic state is observed because the electronic subsystem becomes decoupled from the lattice order due to the strong optical excitation. Indications of the electronic order reacting faster to optical excitation have been observed in multiple experiments\,\cite{Petersen2011, Coslovich2017, Zong2019}; in particular for the case of \tise it has been reported that at very low temperatures lattice order can persist while the electronic order is fully quenched\,\cite{Porer2014a}. In addition recent work has shown that optical excitation leads to a weakening of the electron-phonon coupling strength\,\cite{Otto2021, Heinrich2023} in \ttise.

\section*{Conclusion} 
In conclusion we have shown how intense light pulses can modify the band structure of \tise almost instantly via screening, transitioning it into a non-equilibrium semimetallic state. This result emphasizes how optical excitation can be used as a knob to tune electron-electron and electron-lattice interactions in solids, while driving phases with drastically different properties unreachable in equilibrium.

\section*{Methods}
Time-resolved ARPES measurements were conducted at the Lawrence-Berkeley National Laboratory with 22.3\,eV extreme-ultraviolet (XUV) femtosecond pulses. Photoelectrons are detected with a hemispherical electron analyzer (Scienta R4000). The light source is a cryo-cooled regenerative amplifier (KMLabs Wyvern 500) seeded by the output of a home-built, 76 MHz Ti:sapphire oscillator, which is pumped by 4.5\,W from a green solid-state laser (Lighthouse Photonics Sprout).  The amplifier stage, in turn, is pumped by two green, nanosecond pulsed Nd:YVO$_4$ lasers (Photonics Industries DS20HE). The XUV is created after second harmonic generation by tightly focusing 390\,nm pulses into Kr gas. A detailed description of the setup can be found in \cite{Buss2019}. For the experiment a pump wavelength of 780\,nm with a repetition rate of 25\,kHz was used with a total energy resolution of $\sim$75\,meV and a temporal resolution of about 55\,fs.




\bibliographystyle{ScienceAdvances}
\bibliography{TiSe16}

\section*{Acknowledgments}
This work was primarily funded by the U.S. Department of Energy (DOE), Office of Science, Office of Basic Energy Sciences, Materials Sciences and Engineering Division under contract no. DE-AC02-05CH11231 (Ultrafast Materials Science program KC2203). A.L.\ also acknowledges support from the Gordon and Betty Moore Foundation EPiQS Initiative through Grant No. GBMF4859 for the implementation of the experimental setup. G.M.\ and M.C.\ acknowledge the CINECA award under the ISCRA initiative and PRACE for high performance computing resources. G.M. and M.C. acknowledge support from the European Union (ERC, DELIGHT, 101052708).

\section*{Competing Interest}
The authors declare no competing financial interest.

\section*{Author Contributions}
A.L.\ designed and supervised the project. Data was collected by M.H.\ and Y.L., analyzed by M.H.\ with help from Y.L.\ and A.L.\ and discussed with all authors. Theoretical calculations were performed by G.M.\ and M.C.\
Manuscript preparation was done by M.H.\ with input from all co-authors.


\end{document}